\documentclass[prl,twocolumn,showpacs,floatfix,amsmath,amssymb,superscriptaddress]{revtex4-1}
\usepackage{amsfonts}
\usepackage{stmaryrd}
\usepackage{bbm}
\usepackage{mathrsfs}
\usepackage{tipa}
\usepackage{amssymb}
\usepackage{txfonts}
\usepackage{graphicx}
\usepackage{dcolumn}
\usepackage{epstopdf}
\usepackage[colorlinks,linkcolor=blue,urlcolor=blue,citecolor=blue]{hyperref}
\usepackage{multirow}
\usepackage{subfigure}
\usepackage{url}

\begin{document}
\newcommand*{\cm}{cm$^{-1}$\,}
\newcommand*{\Tc}{T$_c$\,}

\title{Single crystal growth and physical property characterizations of intermediate valent compound YbFe$_2$Al$_{10}$}

\author{J. L. Lv}
\affiliation{Institute of Physics, Chinese Academy of Science, Beijing 100190, China}

\author{R. Y. Chen}
\affiliation{International Center for Quantum Materials, School of Physics, Peking University, Beijing 100871, China}

\author{H. P. Wang}
\affiliation{International Center for Quantum Materials, School of Physics, Peking University, Beijing 100871, China}

\author{J. L. Luo}
\affiliation{Institute of Physics, Chinese Academy of Science, Beijing 100190, China}
\affiliation{Collaborative Innovation Center of Quantum Matter, Beijing 100871, China}

\author{N. L. Wang}
\email{nlwang@pku.edu.cn}
\affiliation{International Center for Quantum Materials, School of Physics, Peking University, Beijing 100871, China}
\affiliation{Collaborative Innovation Center of Quantum Matter, Beijing 100871, China}

\begin{abstract}
We report single crystal growth and physical properties characterization of YbFe$_2$Al$_{10}$ compounds. The measurements of resistivity, magnetic susceptibility, and specific heat show different behaviors from previous studies on polycrystal samples. An intermediate valent characteristic with moderate mass enhancement is indicated. In particular, the optical spectroscopy measurement reveals formation of multiple hybridization energy gaps which become progressively pronounced at low temperature. The multiple hybridization energy gaps are likely caused by the hybridizations between the flat band from Yb 4$f$ electrons and different bands of conduction electrons.
\end{abstract}

\maketitle
\section{Introduction}

The competing and coexisting of different interactions lies in the center of condensed matter physics, which can leads to a wide range of novel quantum properties, such as unconventional superconductivity, non-Fermi liquid behavior, quantum criticality, Kondo semiconductor/insulator etc.\cite{ISI:000300129400001,ISI:A1996UU48300008,ISI:A1976CF50600037,ISI:A1976BR87100018,ISI:A1994NU01300130,ISI:A1994NU01300129,ISI:000167009600007,ISI:000089438700003}.
Especially, for the materials containing rear-earth or actinides elements, the strongly correlated $f$ electrons play an important role and most of them exhibit heavy fermion behaviors.
At relatively high temperatures, the $f$ electrons are essentially localized because there is no direct overlap between their wave functions, and the corresponding electronic band lies well below the Fermi level.
When the temperature decreases, however, the interaction between conduction band and $f$-electrons, usually referred to as Kondo interaction, tends to screen the $f$-electron local moment collectively and forms Kondo singlet state. The Kondo singlet will give rise to a narrow resonance band near the Fermi level, which hybridizes with conduction electron band and leads to a hybridization gap\cite{Abrikosov,PhysRev.138.A515,ISI:000383955500002}. Beyond that, there also exist exchange interactions between those local moments, the Rudermann-Kittel-Kasuya-Yosida (RKKY) interaction, which tend to form long range magnetic orders in the system\cite{PhysRev.96.99,Kasuya01071956,PhysRev.106.893}. The competition between Kondo and RKKY interactions results in a rich phase diagram of heavy fermion materials. The interaction between strongly correlated $f$ electrons and conduction band can be well described by the periodical Anderson model (PAM), based on which the strength of Kondo interaction $\tilde{V}$ can be characterized by the energy scale of the direct hybridization gap $\Delta=2\tilde{V}$. When $\tilde{V}$ is large enough, some of the rear-earth elements will exhibit intermediate valence behavior, that is, the rare earth or actnide elements hold for a non-integer valence.

The orthorhombic ternary compound RT$_2$Al$_{10}$ (R=rear earth elements; T= Fe, Ru, Os)\cite{ISI:000071595100024} have drawn tremendous interest in the community because they have displayed a lot of intriguing properties\cite{ISI:000291190900013,PhysRevB.90.224425,ISI:000208728400021}. For example, an unexpected antiferromagnetic phase transition was identified to occur at 27 K in CeRu$_2$Al$_{10}$ compound\cite{Strydom20092981,PhysRevB.82.100405,ISI:000389503200002}, while CeFe$_2$Al$_{10}$ was reported to be a Kondo insulator\cite{JPC2010,ISI:000323502800058}. The origin of these peculiarities is still under intense debate.
Nevertheless, the Yb-based  RT$_2$Al$_{10}$ seems to be much less investigated, partially due to the difficulty of growing single crystalline samples with good quality.  Among them, YbFe$_2$Al$_{10}$ was found to be an intermediate valence candidate in which Yb shows $+3$ state at high temperatures above 400 K but fluctuates at lower temperatures\cite{ISI:000345658100001,ISI:000316290200021}. However, the Kondo lattice type characters only presented themselves at very low temperatures below 10 K, which is in conflict with the expected strong hybridization strength for intermediate valence compounds.

Up to now, all the measurement on YbFe$_2$Al$_{10}$ were conducted on polycrystalline samples as far as we know, whereas the data of single crystals are still lacking.
In order to unravel the seemingly exotic contradictions, we have successfully synthesize large pieces of YbFe$_2$Al$_{10}$ single crystal using self-flux method, the properties of which are proved to be quite different from that of polycrystalline samples in the literature. We have also performed infrared spectroscopy measurements, which is an ideal method to investigate heavy fermion materials for it can provide direct information on not only the magnitude of the hybridization gap but also the enhancement of effective mass. The results indicate that YbFe$_2$Al$_{10}$ is indeed an intermediate valence compound with quite strong \emph{c-f} hybridization strength.

\section{Crystal growth and characterizations}

\begin{figure}[htbp]
  \centering
  \includegraphics[width=7.5cm]{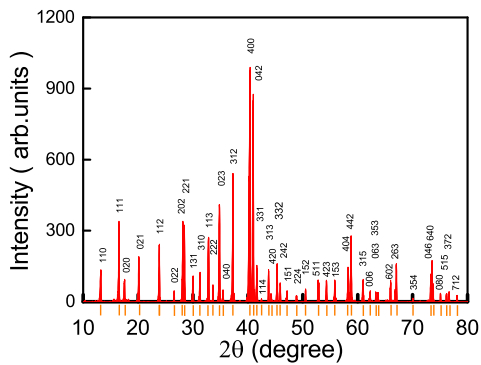}\\
  \caption{(Color online) The powder x-ray diffraction patterns of YbFe$_2$Al$_{10}$ single crystals at room temperature.  }\label{Fig:XRD}
\end{figure}

Single crystal samples of YbFe$_2$Al$_{10}$ are synthesized by self-flux method. High purity (more than 99.99\%) Yb,Fe and Al powder were put in an alumina crucible with a starting composition of YbFe$_2$Al$_{40}$, then sealed in a Ta tube filled with Ar gas.  The Ta tube was then sealed in an evacuated quartz tube, which was heated up to 1200 $^\circ$C in 15 hours, dwelled for 40 h at this temperature and slowly cooled down to 850 $^{\circ}$C at the rate of 2 $^{\circ}$C/h. Large pieces of single crystals with shiny surfaces were yielded by eliminating excess Al flux through a centrifugal machine. The composition of these compounds were confirmed by energy dispersive X-ray (EDX) measurement at different locations in the crystals on different batches gave the average elemental ratio close to 7.75:15.30:76.95, which is close to the nominal stoichiometry within the measurement accuracy. The single crystals were grounded into powders for X-ray diffraction (XRD) measurement. Figure \ref{Fig:XRD} shows the powder XRD pattern. It indicates the pure phase of the grounded crystals. The YbFe$_2$Al$_{10}$ single crystal has the structure with space group Cmcm and the refined lattice constants are a=8.9646(1) \AA, b=10.1282(2) \AA, c=8.9842(4) \AA, which are close to the early report \cite{ISI:000071595100024}.

\begin{figure}[htbp]
  \centering
  \includegraphics[width=7.5cm]{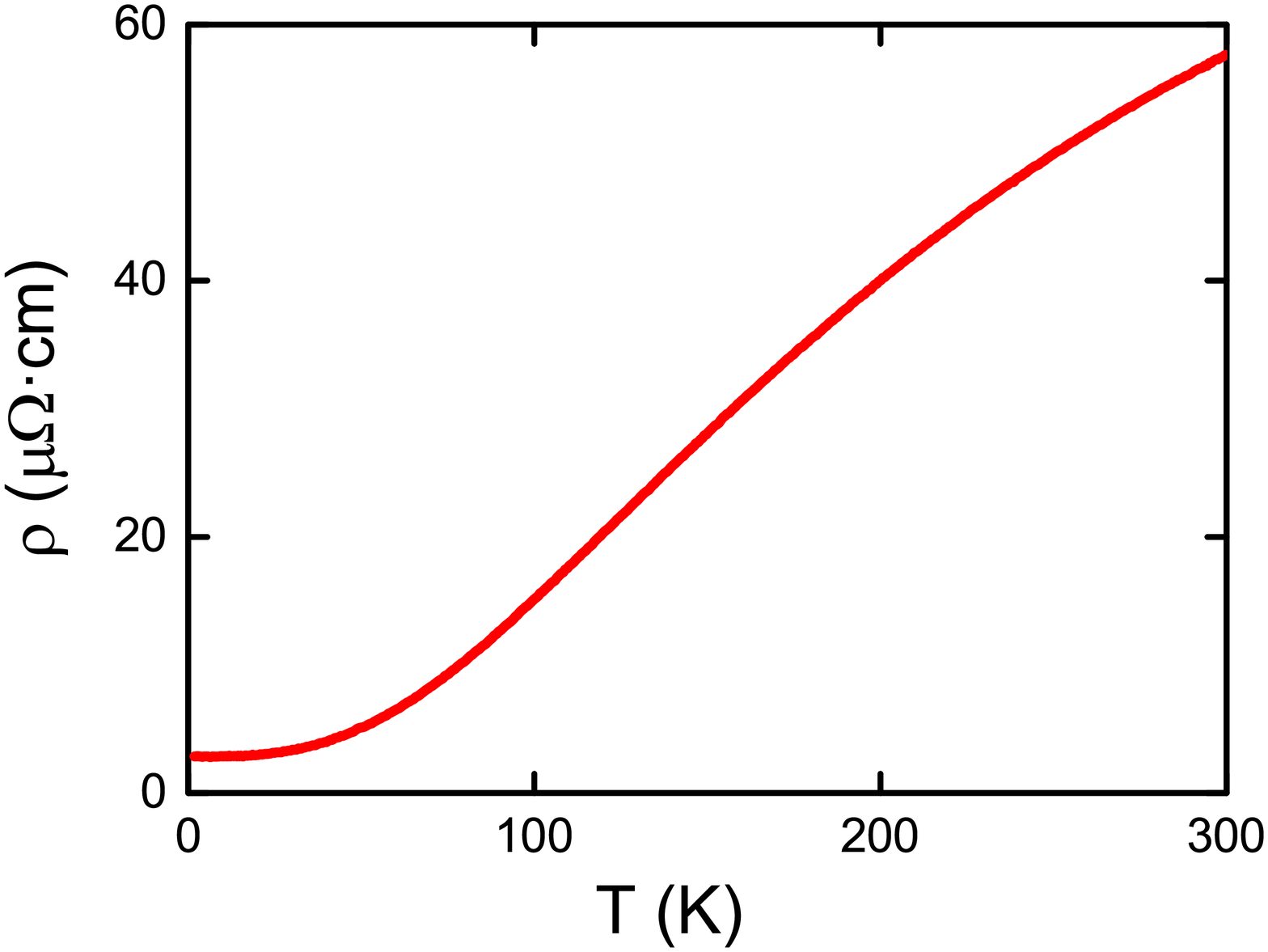}\\
  \caption{The temperature dependent resistivity $\rho$(T) of YbFe$_2$Al$_{10}$ from 1.8 K to 300 K. }\label{Fig:resis}
\end{figure}

The dc resistivity was measured by a four-probe method from room temperature to 1.8 K in a Quantum Design Physical Property Measurement System (PPMS). In a previous report on polycrystal samples, the temperature dependent resistivity $\rho(T)$ of of both YbFe$_2$Al$_{10}$ and YFe$_2$Al$_{10}$ show upturns around 20 K\cite{ISI:000286041300006}. As a contrast, the resistivity of single crystalline  YbFe$_2$Al$_{10}$ follows a good metallic behavior in the temperature range from 2 K to 300 K, as displayed in Fig. \ref{Fig:resis}. The overall  $\rho(T)$  in our measurement is an order of magnitude smaller than that of the polycrystal samples while the residual resistivity ratio (RRR) to that at 300 K is of about 17, both of which indicate that the single crystals are of high quality. Furthermore, $\rho(T)$ of polycrystalline YbFe$_2$Al$_{10}$ shows a peak at 4 K, which was considered as the mark of Kondo screening onset and thus leads to a very small Kondo temperature ($T_K\simeq 1$ K). However, we do not observe any signature of such behavior down to the lowest temperature in our measurement, which immediately raises a question on the energy scale of Kondo interaction in this material.

\begin{figure}[htbp]
  \centering
  \includegraphics[width=7.5cm]{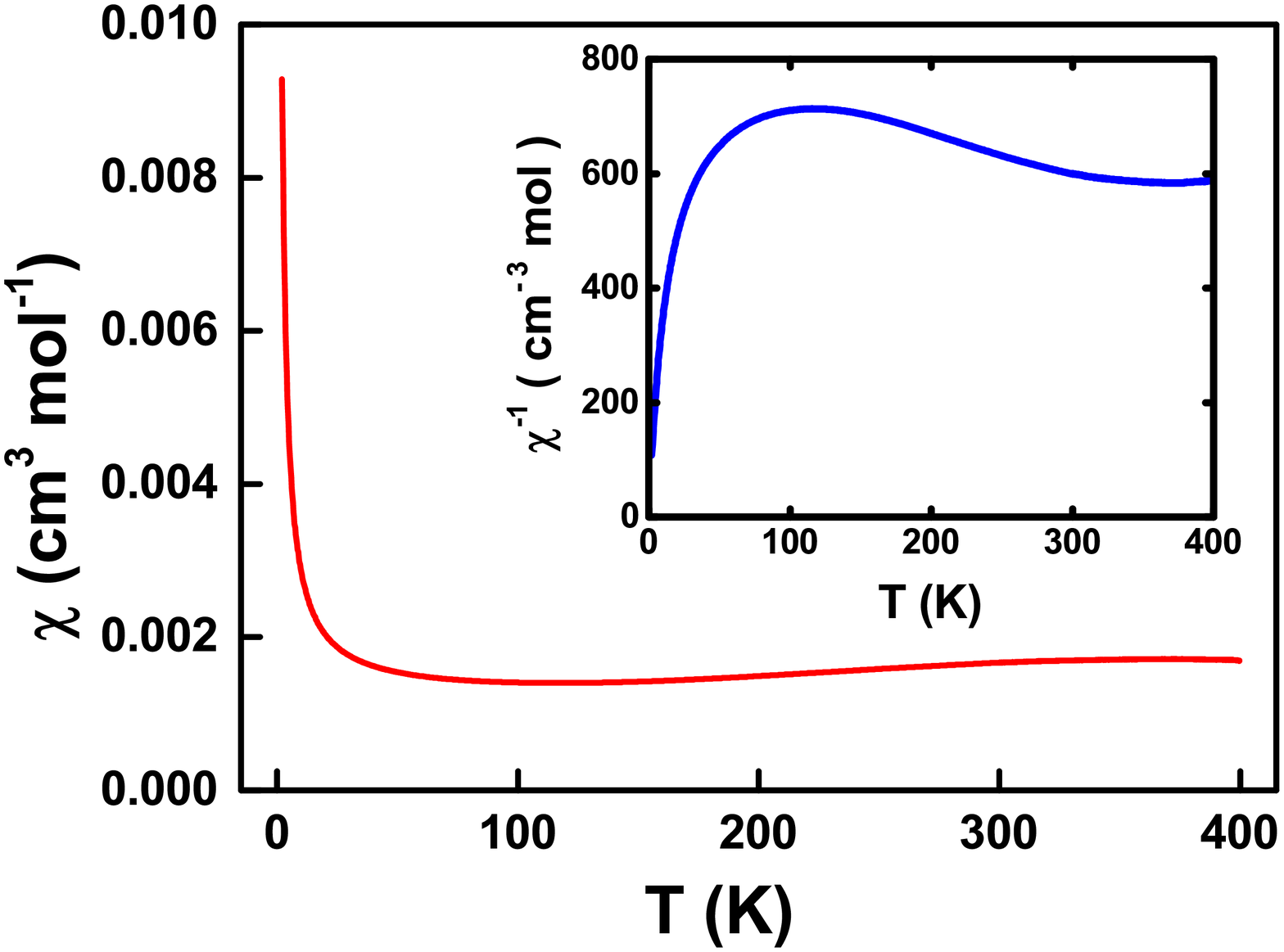}\\
  \caption{Temperature dependent of the susceptibilities from 1.8 K to 400 K measured under a magnetic field of 0.5 T. Inset: a plot of 1/$\chi(T)$ of the YbFe$_2$Al$_{10}$ compound as a function of temperature. }\label{Fig:magnetic}
\end{figure}

The magnetic susceptibility $\chi(T)$ was measured by a Quantum Design Magnetic Property Measurement System (MPMS) under a magnetic field of 0.5 T, as shown in the main panel of Fig. \ref{Fig:magnetic}. Compared with the results of polycrystalline samples, which generally demonstrates two sets of Curie-Weiss behaviors at high temperatures \cite{ISI:000345658100001}, the most prominent feature of $\chi(T)$ is the presence of a broad peak feature at high temperature near 380 K and meanwhile a Curie-Weiss tail at low temperature. The low temperature Curie-Weiss tail could be ascribed to the unscreened magnetic moments or presence of magnetic defects or impurities, which can be fitted by the Curie-Weiss law with a small effective moment of 0.4 $\mu_B$ being derived. The role of 3d-element Fe in this structure-type compounds is still illusive. An early measurement on the isostructural compounds YFe$_2$Al$_{10}$ and LaFe$_2$Al$_{10}$ indicated Pauli paramagnetism, implying the non-magnetic nature of Fe ions \cite{ISI:000071595100024}, while other work on YFe$_2$Al$_{10}$ suggested the existence of weak ferromagnetic correlations \cite{ISI:000286041300006,PhysRevB.86.220401}. We consider that the Fe ions in YbFe$_2$Al$_{10}$ play a minor role of the magnetic property in contrast to Yb ions. Below the broad peak in $\chi(T)$ (or dip in $1/\chi(T)$ in the inset) around 380 K, $\chi(T)$ decreases with decreasing temperature. This behavior is also in contrast to the property of polycrystalline samples where no decreasing feature in $\chi(T)$ was observed \cite{ISI:000345658100001}. Assuming that Yb has a valence state of 3+ and the $\chi(T)$ follows a Curie-Weiss law above 380 K, the broad peak feature could be attributed to the intermediate valence behavior of Yb irons formed at lower temperature as well as the crystal field effect, which suggests that part of the local moments is screened by conduction electrons. In order to confirm this feature, further studies including magnetic susceptibility data higher than 400 K are needed.

\begin{figure}[htbp]
  \centering
  \includegraphics[width=7.5cm]{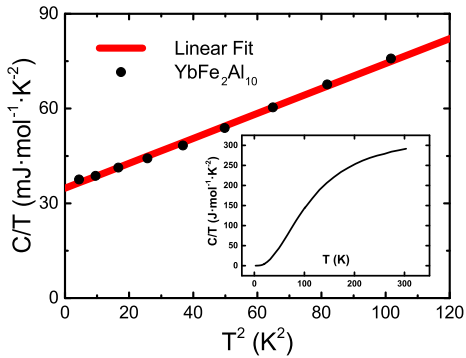}\\
  \caption{The temperature dependent specific heat $C_p(T)$ of YbFe$_2$Al$_{10}$ single crystal. The main panel is a plot of  C$_p$/T as a function of $T^2$ together with a fit to a straight line at the low temperature. Inset is a plot of specific heat data from 2 K to 300 K. }\label{Fig:heat}
\end{figure}

The specific heat $C_p(T)$ of YbFe$_2$Al$_{10}$ single crystal measured in a Quantum Design Physical Property Measurement System (PPMS)  is shown in Fig. \ref{Fig:heat}. In order to decompose the contributions from different components, we fit the low temperature $C_p(T)$ with a formula $C_p$/T =$\gamma$ + $\beta$ $T^2$, where $\gamma$ is the Sommerfeld coefficient. The first and second terms represent the contributions from itinerant carriers and phonons, respectively. The obtained Sommerfeld coefficient is about 35 $mJK^{-2} mol^{-1}$, much smaller than that of the prototype heavy fermion materials like CeCoIn$_5$. The moderate value of $\gamma$ implies a mild effective mass enhancement, which is correlated with a high temperature scale at which the magnetic susceptibility deviates from the Curie-Weiss behavior. We notice that the Sommerfeld coefficient of many intermediate valence compounds are comparable to YbFe$_2$Al$_{10}$,  such as YbAl$_3$ \cite{Ebihara2000754}.

\section{Optical spectroscopy}

\begin{figure}[htbp]
\centering
\includegraphics[width=7.5cm]{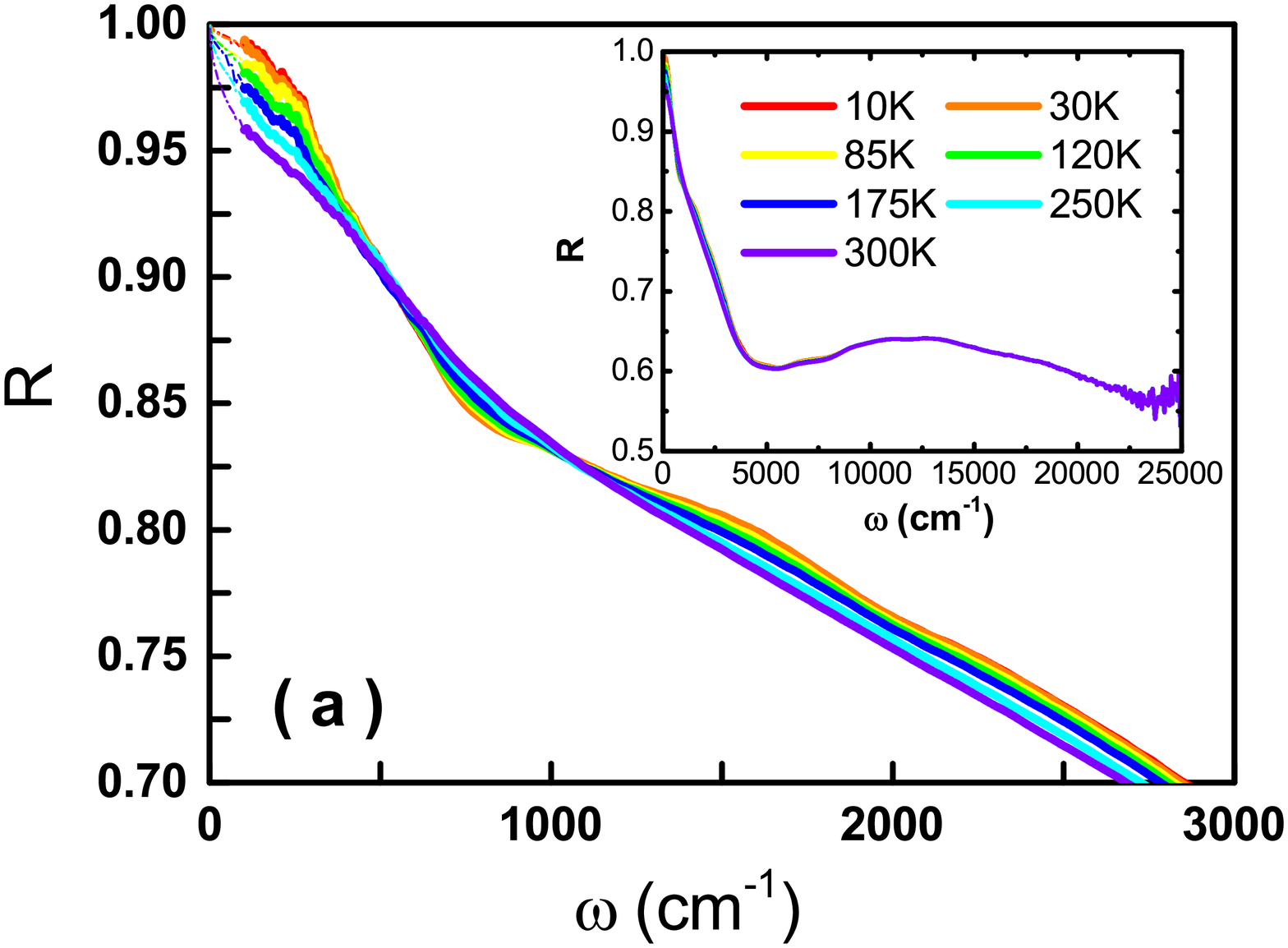}\vspace{-5pt}
\includegraphics[width=7.5cm]{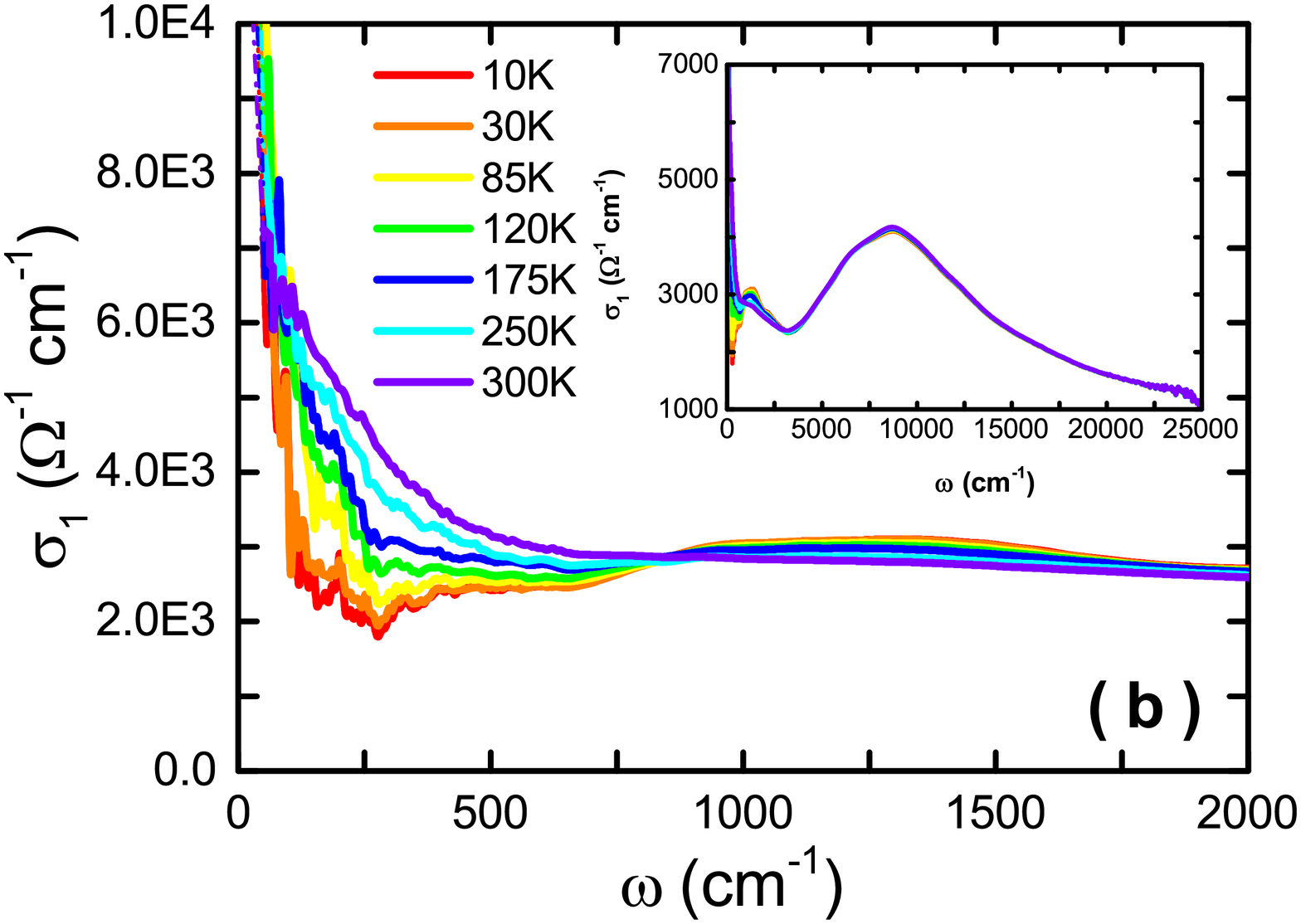}
\caption{(a) Optical reflectance spectra of YbFe$_2$Al$_{10}$ single crystal at different temperature points. (b) Optical conductivity spectra $\sigma_1(\omega)$ obtained from reflectance spectra using Kramers-Kronig transformation. Insets are spectra over broad frequency ranges.} \label{fig:ref}
\end{figure}

We measured the temperature-dependent optical reflectance on Bruker 113v and Vertex 80v spectrometers on shinny as-grown surface of YbFe$_2$Al$_{10}$ single crystals by an \emph{in-situ} gold and aluminum overcoating technique from 40 to 25000 \cm. The real part of the conductivity $\sigma_1(\omega)$ was derived by the Kramers-Kronig transformation of $R(\omega)$. The optical reflectivity $R(\omega)$ and conductivity $\sigma_1(\omega)$ at several selective temperatures are displayed in Fig. \ref{fig:ref} (a) and (b), respectively. The reflectivity $R(\omega)$ approaches to unit at zero frequency for all the measured temperatures and the low energy part increases with temperature decreasing, both of which are typical metallic responses. At room temperature, $R(\omega)$ increases monotonically as frequency decreasing, whereas a broad dip feature gradually emerges around 800 \cm at lower temperatures, which gets more pronounced upon cooling. Such depletion is commonly observed in heavy fermion materials, originating from the formation of hybridization energy gap in the density of states near the Fermi level. In addition, a double peak structure was identified at higher energies around 1200 \cm and 2400 \cm respectively.

The physical properties are more clearly revealed by the optical conductivity, as shown in Fig. \ref{fig:ref} (b). At room temperature, the low energy $\sigma_1(\omega)$ is characterized by two Drude peaks centered at zero frequency, representing the responses of itinerant carriers. Corresponding to the reflectivity, two very weak bumps were also observed locating around 1200 \cm and 2400 \cm. With temperature decreasing, both of the two Drude peaks narrow considerably and the spectral weights are substantially reduced. The narrowing of Drude peaks indicates that the scattering rate of free carriers is reduced, which suggests that the strong spin-flip scattering caused by local $f$ moments at higher temperatures is progressively screened by the conduction electrons. The spectral weight (SW) of Drude component is proportional to $n/m^*$, where $n$ and $m^*$ stand for the density and effective mass of the free carriers. Therefore, the reduction of $SW$ infers either the deceasing of free carrier density or the increasing of effective mass. In the Kondo screening regime, since the local f-electrons gradually become itinerant and contribute to the Fermi surface, the carrier density $n$ should increase with cooling, this means that the effective mass was effectively enhanced. Furthermore, a weak peak gradually developed around 440 \cm in the optical conductivity, which is a strong evidence of the opening of an energy gap arising from the hybridization of conduction electrons and Yb $f$ electronic bands. Besides, the two additional weak bumps observed at 300 K persisted in the whole measurement temperature range and become even more prominent upon cooling.

\begin{table*}[htbp]
\setlength\abovecaptionskip{0.5pt}
\caption{Fitting parameters of $\sigma_1(\omega)$ from 10 K to 300 K. At 10 K, an additional Lorentz term centered at 440 \cm is added at 10 k. The unit of all the fitting parameters are \cm. \label{2}}
\vspace{-1em}
\begin{center}
\renewcommand\arraystretch{1.5}
\begin{tabular}{p{1.5cm} p{1.2cm} p{1.1cm} p{1.1cm} p{1.1cm} p{1.1cm} p{1.1cm}p{1.1cm} p{1.1cm} p{1.1cm} p{1.1cm} p{1.1cm} p{1.1cm}p{1cm}}
\hline
\hline
 T&$\omega_{p1}$&$\gamma_{D1}$&$\omega_{p2}$&$\gamma_{D2}$&$S_1$&$\gamma_1$&$\omega_1$&$S_2$&$\gamma_2$&$\omega_2$&$S_3$&$\gamma_3$&$\omega_3$\\
\hline
 10 K&7910&25&5280&206&4900&390&442&17370&1880&1220&5650&1300&2410\\
 30 K&8040&26&6270&256&4340&390&442&17280&1900&1220&5566&1300&2410\\
 85 K&8550&30&7010&280&4110&390&440&17190&1920&1210&5510&1300&2410\\
120 K&9110&32&7860&340&3100&390&440&16930&1950&1200&5480&1350&2410\\
170 K&9200&36&8300&370&2480&390&440&16910&2140&1200&5440&1380&2410\\
250 K&9750&40&8970&380&1900&390&437&16800&2130&1200&5420&1400&2410\\
300 K&10490&42&9750&497&-&-&-&17100&2400&1190&5320&1400&2410
\\
\hline
\hline
\end{tabular}\\
\end{center}
\end{table*}
In order to investigate the evolution of physical properties qualitatively, we use the Drude-Lorentz mode to decompose the optical conductivity spectra. The overall dielectric function can be expressed as:
\begin{equation}\label{Eq:DL}
\epsilon(\omega)= \epsilon_{\infty}-\sum_{s}{\frac{\omega_{ps}^2}{\omega^2+i\omega/\tau_{Ds}}}+ \sum_{j}{\frac{S_j^2 }{\omega_j^2-\omega^2-i\omega/\tau_j}}. 
\end{equation}
where $\varepsilon_{\infty}$ is the dielectric constant at high energy; the second and the last terms represent the Drude and Lorentz components, describing the contribution from conduction electrons and interband transitions respectively \cite{ISI:000080050500005}.

 \begin{figure}[htbp]
\centering
\includegraphics[width=7.5cm]{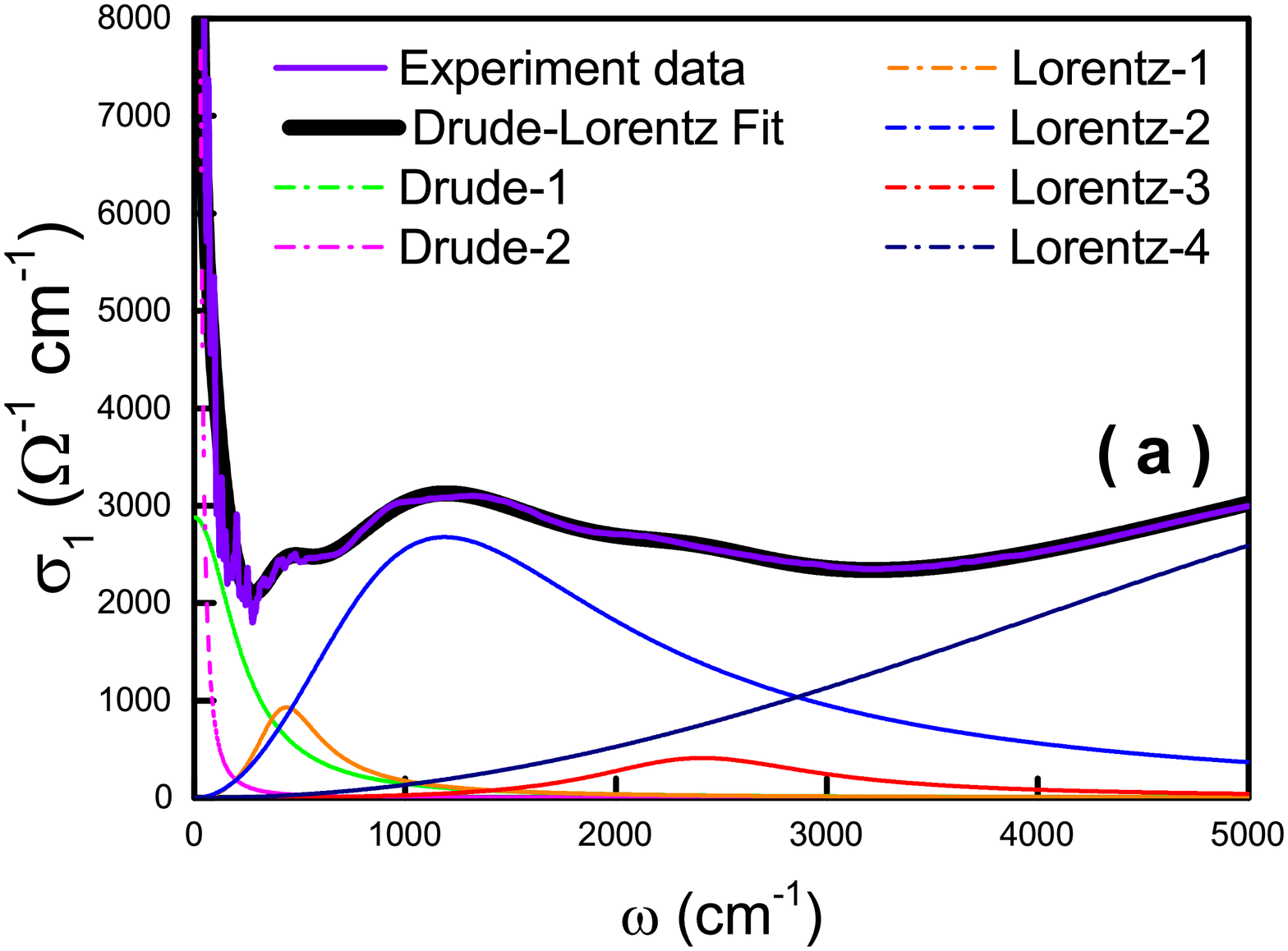}\vspace{-6pt}
\includegraphics[width=7.5cm]{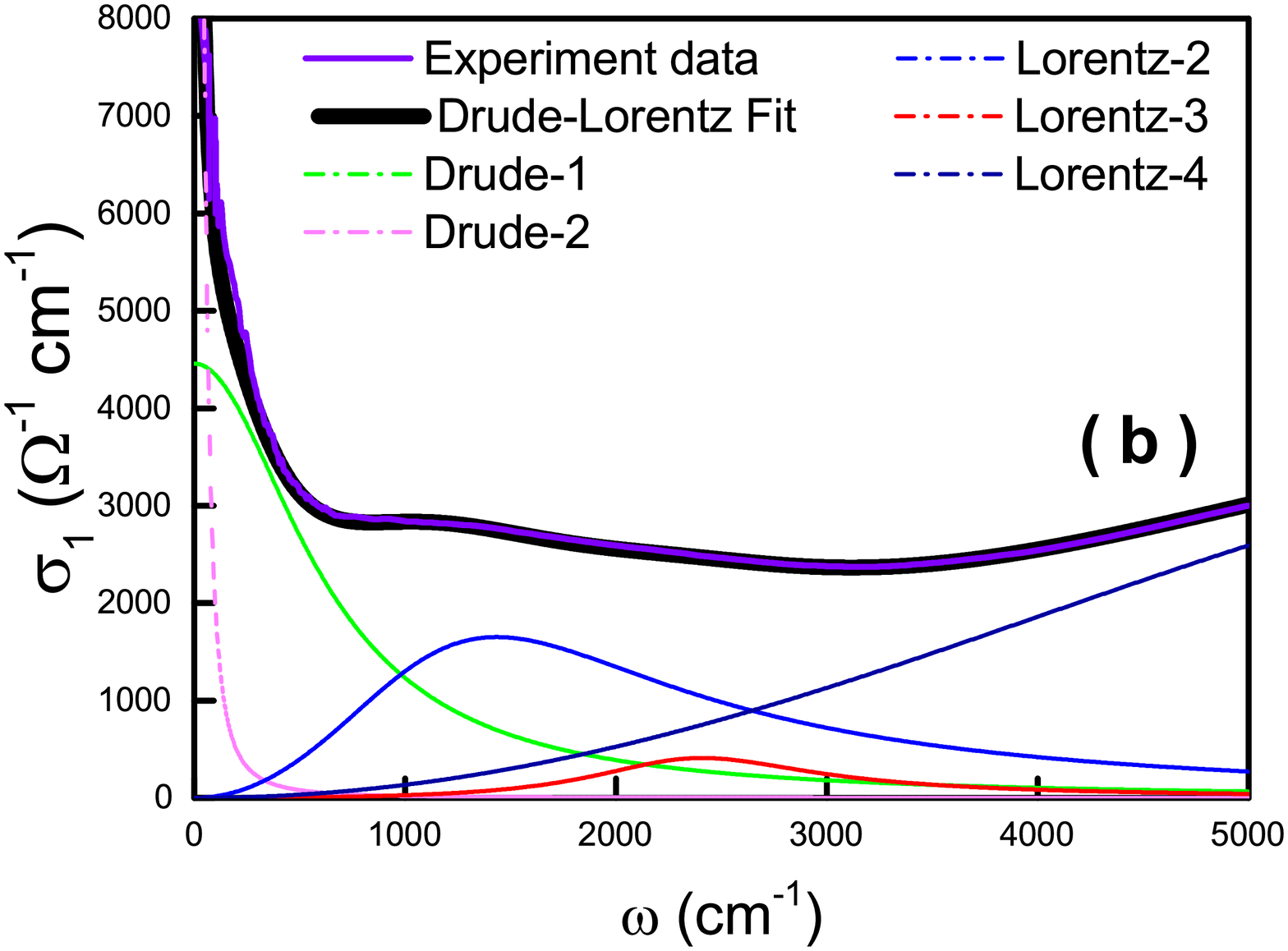}
\caption{The experimental optical conductivity $\sigma(\omega)$ together with fits to the Drude and Lorentz model of YbFe$_2$Al$_{10}$ at (a)10 K and (b) 300 K, repectively. } \label{fig:DLfit}
\end{figure}

\begin{figure}[htbp]
  \centering
  \includegraphics[width=7.5cm]{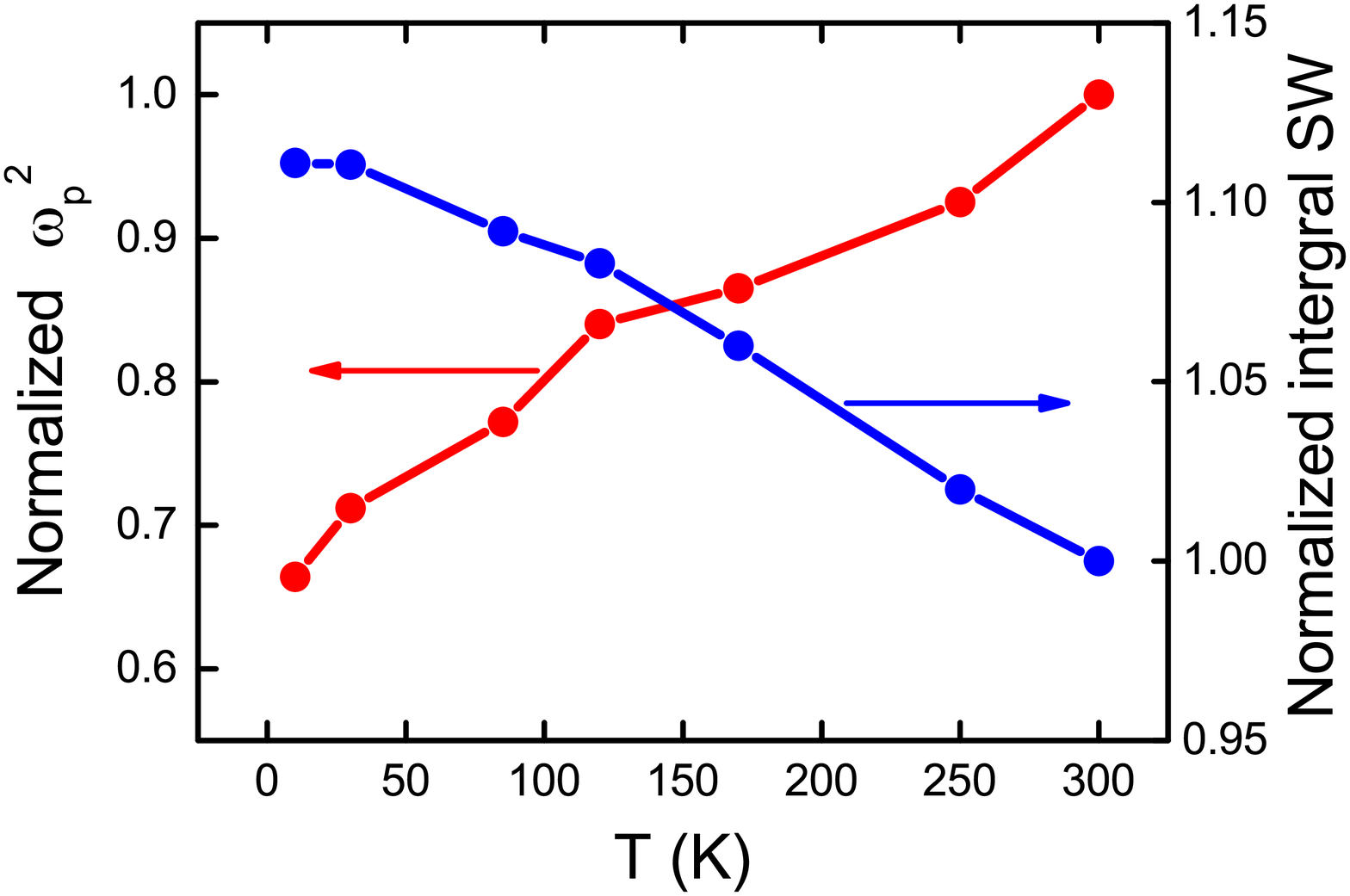}\\
  \caption{Temperature dependent square of plasma frequency $\omega^2_p$ normalized to the value at 300 K is plotted in red curve on left axis. The blue line shows the intergral of the spectra weight over the frequency from 270 \cm at which frequency the relative minimum of $\sigma_1$ located until the frequency the spectra of various T overlapped together.  }
  \label{Fig:SW}
\end{figure}

 As shown in Fig. \ref{fig:DLfit}, the optical conductivity $\sigma_1(\omega)$ can be well reproduced by two Drude terms and several Lorentz components. The requirement of more than one Drude component to reproduce the low energy free carrier contribution suggests presence of multiple bands crossing the Fermi level for YbFe$_2$Al$_{10}$. It is considered $T_K$ to be from averaging over that Fermi surface and hybridization strength can be different around the Fermi surface. As a matter of fact, the mutiple-Lorentz fitting indicats the appearance of several hybridization gaps in the optical conductivity, suggesting different hybridization strength. Part of the obtained fitting parameters at different temperatures are listed in Table \ref{2}. Among them, $\gamma_D=1/\tau_D$ represents the scattering rate of the itinerant carriers, where $\tau_D$ is the life time.  At room temperature, the values of scattering rates are relatively large, $\gamma_{D1}$= 47 \cm and $\gamma_{D2}$=497 \cm, due to the strong scattering of the local moments. With temperature cooling, $\gamma_{D1}$ decreases from 47 \cm to 25 \cm, and $\gamma_{D2}$ drops sharply from 497 \cm to 206 \cm at 10 K. In the meantime, the plasma frequencies $\omega_{p1}$ and $\omega_{p2}$ also experience continuous drops upon temperature cooling, as depicted in Fig. \ref{Fig:SW} by red solid dots. Here, the overall plasma frequency is obtained by $\omega_p^2=\omega_{p1}^2+\omega_{p2}^2$, which decreases from 14300 \cm at room temperature to 9500 \cm at 10 K. The decrease reflects the mass enhancement due to enhanced Kondo screening effect at low temperature. According to the Fermi-liquid theory, $\gamma$ is proportional to n$^{1/3}$m$^*$. Combined the $\gamma$ value from specific heat measurement with the plasma frequency $\omega_p^2$ = $4{\pi} n e^2/m^*$ \cite{PhysRevB.56.1366}, we can obtain the values of the carrier density n$\approx$ 4.5$\times$10$^{22}$ cm$^{-3}$ and effective mass $m^*$ $\approx$ 45 $m_e$ for temperature at 10 K, which are common values in Yb-based intermediate valence compounds as YbInCu$_4$\cite{FIGUEROA1998347}.

Now let us turn to the analysis of the Lorentz components. At room temperature, four Lorentz components are employed to fit the optical conductivity. Two of them represent conventional interband transitions in high energies (above 5000 \cm). Since they are temperature independent, we shall not discuss them. We notice that the two peaks centered at 1200 \cm and 2400 \cm shift slightly to higher energies with temperature decreasing, along with the enhancement of the peak strength, indicating that these two peaks arise from excitations across temperature dependent energy gaps. Furthermore, an additional Lorentz component centering around 440 \cm is required to account for the shoulder feature appearing at lower temperatures, as plotted in Fig. \ref{fig:DLfit} (b). They are most likely to be caused by the \emph{c-f} hybridizations as elaborated above.

Similar observation was reported by Okamura et al. in another intermediate valence compound YbAl$_3$\cite{YbAl3optical,Okamura2004}, where the multiple peaks structure was speculated to arise from the indirect and direct interband transitions across the hybridization gap, respectively. Since the indirect transition could occur only through the assistance of impurities and bosonic excitations (e.g. phonons), the spectral weight associated with the transition is in fact extremely small. Here, we propose that all the three Lorentz peaks observed at 10 K are originated from direct interband transitions across hybridization gaps. Since more than one Drude component have to be used to reproduce the low frequency conductivity spectra, we believe that there should be multiple conduction bands crossing the Fermi level for YbFe$_2$Al$_{10}$. The Kondo interaction could mix the flat band from Yb 4$f$ electrons with different dispersive bands of conduction electrons, giving rise to hybridization gaps of different energy scales. In fact, multiple hybridization energy gap features were observed in CeMIn$_5$ (M=Co, Ir) compounds and well explained by the hybridizations between Ce 4f band and bands from In atoms in different layers in the structure based on calculations from dynamical mean field theory in combination
with local density approximation \cite{Mena2005,Burch2007,Shim2007}. The presence of high energy scales of the gaps is in accord with the expected strong hybridization strength in an intermediate valence compound.

\begin{figure}
\centering
\subfigure{\includegraphics[width=7.5cm]{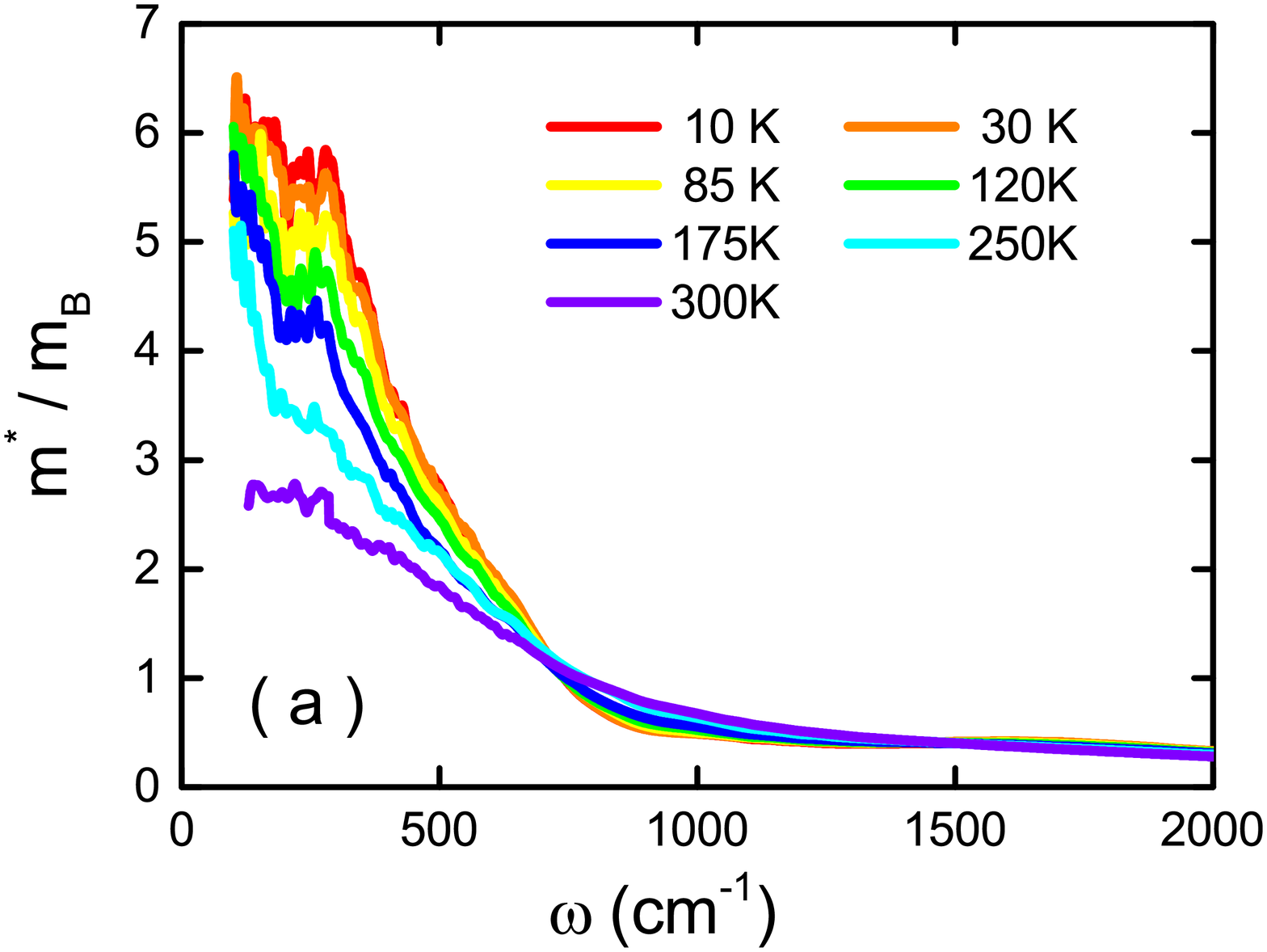}}
\subfigure{\includegraphics[width=7.5cm]{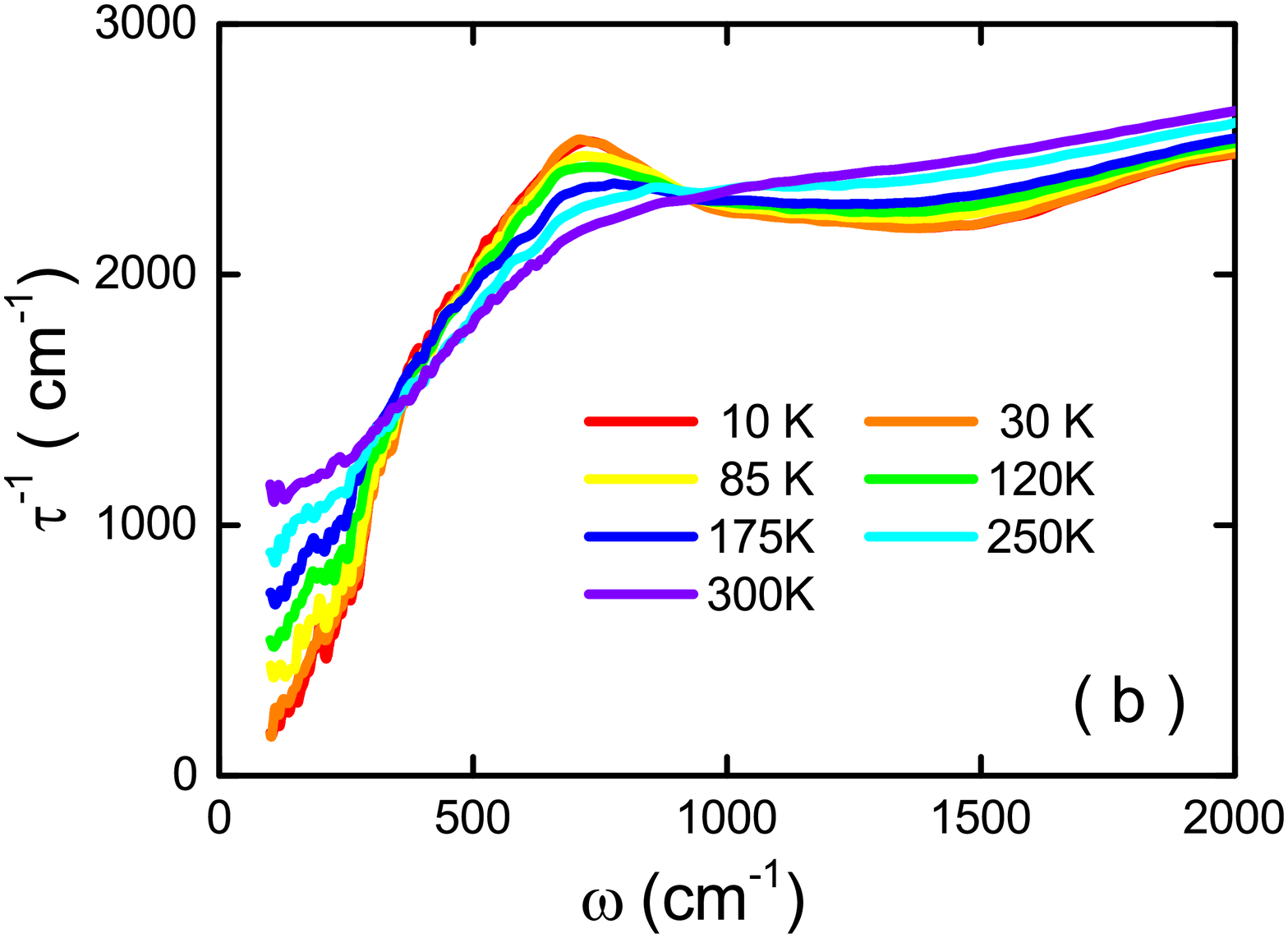}}
\caption{(Color online) (a) Frequency dependence of the effective mass and (b) scattering rate of YbFe$_2$Al$_{10}$.} \label{fig:8}
\end{figure}

Before conclusion, we would like to remark that, for correlated metals including heavy fermion systems and intermetallic compounds, a generalized Drude model is often used in the analysis of infrared conductivity where the simple Drude model is extended by making the damping term complex and frequency as well as temperature dependent, although it does not appear to be an proper approach to systems with presence of multiple conduction bands. According to generalized Drude model, the scattering rate and effective mass could be obtained from the real and imaginary parts of optical conductivity \cite{ISI:000231523800007}:
\begin{equation}\label{Eq:DL}
\frac{1}{\tau(\omega)}= \frac{\omega_{p}^2 }{4 \pi} \frac{\sigma_1(\omega) }{\sigma_1^2(\omega)+\sigma_2^2(\omega)},
\end{equation}
\begin{equation}\label{Eq:DL}
\frac{m^{*}}{m_{B}}= \frac{\omega_{p}^2 }{4 \pi \omega} \frac{\sigma_2(\omega) }{\sigma_1^2(\omega)+\sigma_2^2(\omega)}, 
\end{equation}
where the $\omega_{p}$ is plasma frequency which should include the spectral weight contributed from free carriers. Here we also present the spectral features for scattering rate and effective mass of YbFe$_2$Al$_{10}$ based on this approach. As plotted in Fig. \ref{fig:8}, the scattering rate is suppressed below about 750 \cm and at the same energy position the effective mass strongly enhanced. Such features have been seen in other Kondo lattice systems\cite{ISI:000166539900032}. According to this plot, further increase of the effective mass should be present in the low-frequency extrapolated region.

\section{Summary}

In summary, we have successfully grown high quality single crystal samples of YbFe$_2$Al$_{10}$. The measurements of resistivity, magnetic susceptibility and specific heat exhibit much differences from previous studies on polycrystal samples. Significantly, the magnetic susceptibility shows a broad peak around 380 K, indicating Kondo coherence behavior at lower temperatures. Optical conductivity revealed three energy gaps locating around 440 \cm, 1200 \cm and 2400 \cm respectively, all of which are proposed to be originated from the \emph{c-f} hybridizations, but locating at different positions in the momentum space.  Meanwhile, the scattering rate of free carriers, represented by the half width of the Drude peaks, gets much smaller as temperature cooling, due to the screening of the $f$ local moments. The Drude spectral weight decreased with temperature as well, demonstrating the enhancement of the effective mass of free carriers. Our results not only refresh the basic understandings of the complex intermediate valence compound YbFe$_2$Al$_{10}$, but also provide new insight into other analogous materials.

\begin{center}
\small{\textbf{ACKNOWLEDGMENTS}}
\end{center}

This work was supported by the National Science Foundation of China (No. 11327806, GZ1123), and the National Key Research and Development Program of China (No.2016YFA0300902).

\bibliographystyle{apsrev4-1}
  \bibliography{YbFe2Al10}

\end{document}